\documentclass[aip,jcp,amsmath,amssymb,reprint]{revtex4-2}

\usepackage{graphicx}
\usepackage{dcolumn}
\usepackage{bm}
\usepackage{float}

\DeclareMathOperator\erfc{erfc}

\usepackage[utf8]{inputenc}
\usepackage[T1]{fontenc}
\usepackage{mathptmx}
\usepackage{etoolbox}

\usepackage{upgreek}

\usepackage{listings}

\usepackage[dvipsnames]{xcolor}

\usepackage{lmodern}

\usepackage{cprotect}

\usepackage[hidelinks]{hyperref}

\begin{document}
\definecolor{mygray}{rgb}{0.95, 0.95, 0.95}
\definecolor{nearlywhite}{rgb}{0.995, 0.995, 0.995}
\newcommand{\mywidth}{0.97\linewidth}

\title{Fast, Modular, and Differentiable Framework for Machine Learning-Enhanced Molecular Simulations}
\author{Henrik Christiansen}
\email{henrik.christiansen@neclab.eu}
\author{Takashi Maruyama}
\author{Federico Errica}
\author{Viktor Zaverkin}
\author{Makoto Takamoto}
\author{Francesco Alesiani}
\affiliation{NEC Laboratories Europe GmbH, Kurfürsten-Anlage 36, 69115 Heidelberg, Germany}

\date{\today}

\lstset{
language=Python,
basicstyle=\ttfamily\scriptsize,
keywordstyle = \color{ForestGreen}, 
morekeywords = {dimos,},
stringstyle=\color{red},
commentstyle=\color{blue},
showstringspaces=false,
frame=no,
numbers=none,
numberstyle=\tiny,
breaklines=true,
captionpos=b,
backgroundcolor = \color{mygray}, 
frameshape={RYR}{Y}{Y}{RYR}
}

\begin{abstract}
    We present an end-to-end differentiable molecular simulation framework (DIMOS) for molecular dynamics and Monte Carlo simulations.
    DIMOS easily integrates machine-learning-based interatomic potentials and implements classical force fields including an efficient implementation of particle-mesh Ewald.
    Thanks to its modularity, both classical and machine-learning-based approaches can be easily combined into a hybrid description of the system (ML/MM).
    By supporting key molecular dynamics features such as efficient neighborlists and constraint algorithms for larger time steps, the framework makes steps in bridging the gap between hand-optimized simulation engines and the flexibility of a \verb|PyTorch| implementation.
    We show that due to improved linear instead of quadratic scaling as function of system size DIMOS is able to obtain speed-up factors of up to $170\times$ for classical force field simulations against another fully differentiable simulation framework.
    The advantage of differentiability is demonstrated by an end-to-end optimization of the proposal distribution in a Markov Chain Monte Carlo simulation based on Hamiltonian Monte Carlo (HMC).
    Using these optimized simulation parameters a $3\times$ acceleration is observed in comparison to ad-hoc chosen simulation parameters.
    The code is available at \url{https://github.com/nec-research/DIMOS}.
\end{abstract}

\maketitle
\section{Introduction}
Molecular simulations are a cornerstone of modern computational physics, chemistry and biology, enabling researchers to understand complex properties of systems.\cite{hollingsworth2018molecular}
Traditional molecular dynamics (MD) and Markov Chain Monte Carlo (MCMC) simulations rely on pre-defined force fields and specialized software\cite{LAMMPS,eastman2023openmm} to achieve large timescales and efficient sampling of rugged free-energy landscapes.\cite{janke2007rugged}
These simulation packages generally lack the flexibility and modularity to easily incorporate cutting-edge computational techniques such as machine learning (ML) based enhancements without major changes to the code base:
For example, the simulation engine LAMMPS\cite{LAMMPS} by now supports certain popular machine learning interatomic potentials (MLIPs)\cite{friederich2021machine} such as MACE,\cite{Batatia2022mace} but novel architectures or new atomic descriptors require adaptation both of the model itself and the core simulation engine.
Other ML based approaches relying on the end-to-end differentiability of the simulation framework are not at all supported by classical simulation packages.
On the other hand, the design principle of ML focused packages is often prioritizing usablity over performance. 
\par
Here, we present an end-to-end differentiable molecular simulation framework (DIMOS) implemented in \verb|PyTorch|,\cite{NEURIPS2019_9015} a popular library for ML research.
Our design goal was a fast simulation engine in a high-level languange with ML based approaches at it's core.
For this, DIMOS implements essential algorithms to perform MD and MCMC simulations, providing an easy-to-use way to interface MLIPs and an efficient implementation of classical force field components in addition to common integrators and barostats.
This is complimented by an implementation of efficient calculation of neighborlists and constraint algorithms which allow for larger timesteps of the numerical integrator.
By relying on \verb|PyTorch|, we inherit many advances achieved by the ML community: Fast execution speed on diverse hardware platforms, combined with a simple-to-use and modular interface implemented in \verb|Python|.
\par
DIMOS is designed as a complementary engine for fast, differentiable prototyping and method development rather than a direct replacement for classical optimized MD codes, trading some peak performance for flexibility and end-to-end differentiability.
The package fills an important void in the landscape of existing differentiable simulation frameworks\cite{jaxmd2020,doerr2021torchmd,Greener2024,ple2024fennol,cohen2025torchsimefficientatomisticsimulation} which currently suffer from respective limitations.
For example the two \verb|PyTorch| based packages we are aware of, torchMD\cite{doerr2021torchmd} and TorchSim,\cite{cohen2025torchsimefficientatomisticsimulation} either do not implement neighborlists leading to computationally prohibitive simulation times or do not support setting up complex classical force field simulations.
Note that TorchSim\cite{cohen2025torchsimefficientatomisticsimulation} was released after DIMOS.
The other packages are implemented in either Jax or Julia: Jax M.D.\cite{jaxmd2020} and FeNNol\cite{ple2024fennol} are Jax\cite{jax2018github} based, and Molly.jl\cite{Greener2024} is implemented in Julia.\cite{Julia-2017}
As consequence, the integration of \verb|PyTorch| based code or models is not straightforward.
Generally we want to note that while there exist separate implementations for many of the algorithms of DIMOS in various code bases, we are not aware of a unified simulation framework implementing them as comprehensively.\footnote{For example, see https://github.com/openmm/NNPOps for implementations of neighborlists and particle-mesh Ewald methods.}
\par
Automatic differentiation, a cornerstone of ML, opens the door for novel approaches by enabling the automatic calculation of derivatives of arbitrary components in the computation graph.
For example, differentiating through the simulation either to tune force-field parameters based on desired dynamic properties\cite{goodrich2021designing} or to adapt the parameters of simulation methods to accelerate the sampling in equilibrium\cite{christiansen2023self} becomes possible, representing only a small subset of potential future developments.
\par
In the remainder of the manuscript, we will discuss the components implemented in DIMOS, and how they can be interfaced in Section~\ref{sec:methods}.
Section~\ref{sec:benchmark} presents benchmark results for water boxes highlighting the scaling behavior and the performance for protein systems is presented, discussing the use of hybrid classical and ML based modelling of interactions.
The end-to-end differentiability is demonstrated in Section~\ref{sec:hmc} by tuning parameters of HMC for alanine-dipeptide in explicit solvent.
Finally, we conclude in Section~\ref{sec:conclusion}.

\begin{figure*}
  \begin{center}
      \begin{minipage}{0.3\textwidth}\centering \includegraphics[width=\textwidth]{{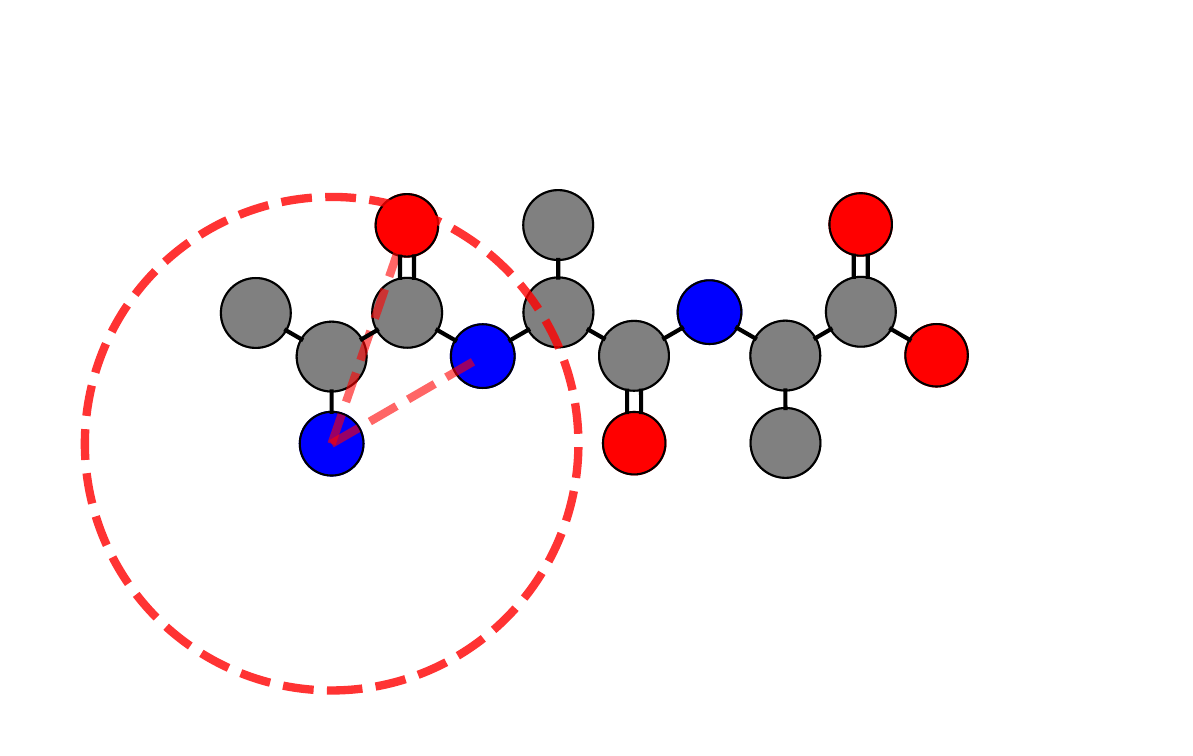}} (a) Classical (MM) \end{minipage}
      \begin{minipage}{0.3\textwidth}\centering \includegraphics[width=\textwidth]{{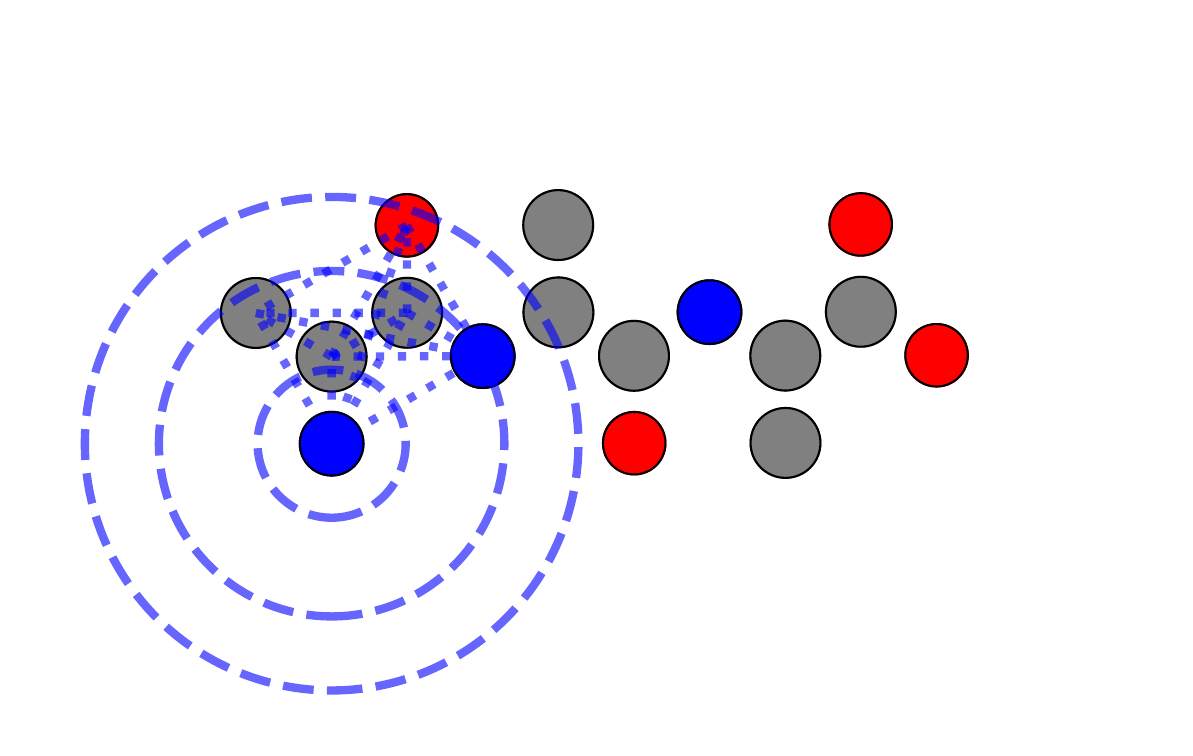}} (b) MLIP (ML)\end{minipage}
      \begin{minipage}{0.3\textwidth}\centering \includegraphics[width=\textwidth]{{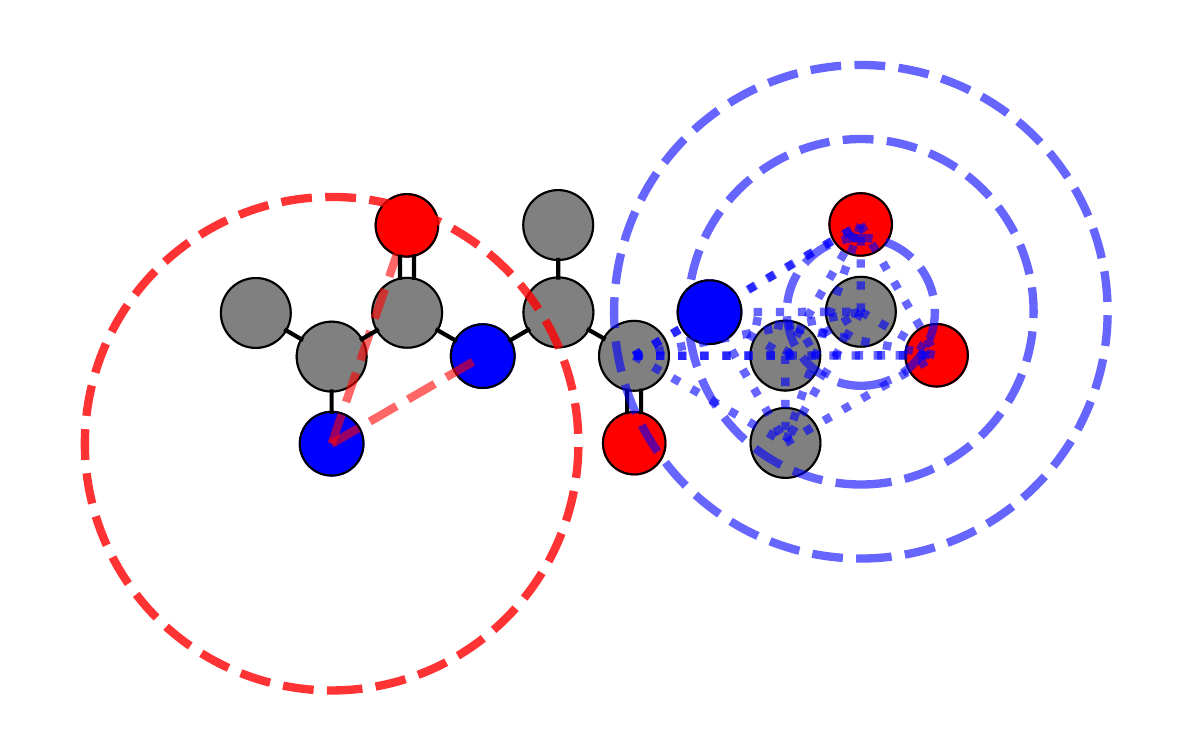}} (c) Hybrid approach (ML/MM) \end{minipage}\\\vspace{0.5cm}
      \begin{minipage}{0.9\textwidth}\includegraphics[width=\textwidth]{{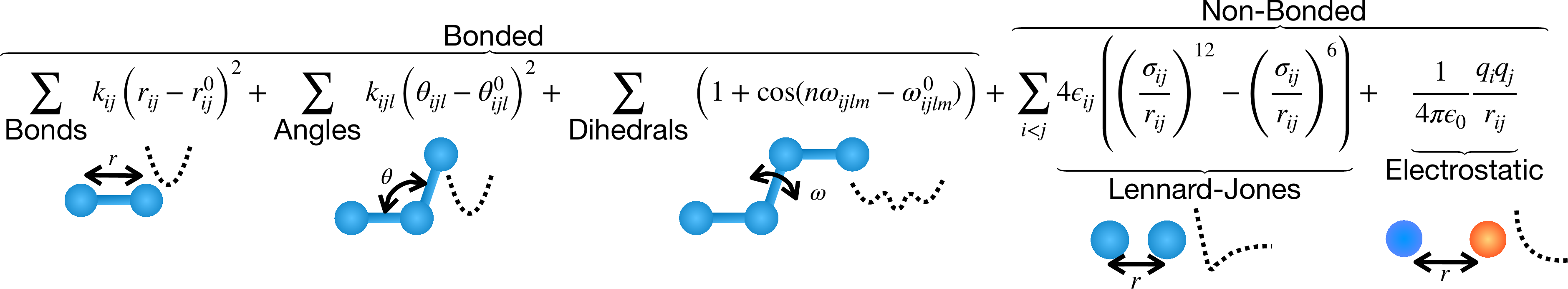}} (d) Typical energy components for a classical description of interactions based on the AMBER force field.\cite{weiner1981amber}\end{minipage}
  \end{center}
  \caption{In (a)-(c) a schematic representation of Alanine tripeptide (Ala$_3$), where the atomic interactions are modeled (a) solely classically (MM), (b) solely using an MLIP (ML), or (c) using a hybrid approach (ML/MM). The red circle marks a region modeled by a classical force field, where one defines interaction terms based on covalent bonds and non-covalently bonded atoms. The blue circles represent the message passing iterations of a graph neural network used to construct many-body interactions, for details see Ref.~\onlinecite{duval2023hitchhiker} and the references therein. In all cases, only the operations based on calculating the force/energy for a single atom are depicted, and for a full evaluation the steps are repeated for every atom. In (d), the energy components for a classical force field are depicted, based on the parameterization of AMBER class interactions.\cite{weiner1981amber} Covalently bonded and non-bonded interactions are distinguished, with exclusions in the non-bonded interactions for directly or indirectly bonded atoms. For more details on the individual contributions, see Section~\protect\ref{sec:FF}.}
  \label{fig:schematic}
\end{figure*}
\section{Simulation Package}
\label{sec:methods}
Setting up a minimal simulation based on classical force fields in DIMOS can be done in only a few lines:
\begin{center}
\begin{minipage}{\mywidth}
\begin{lstlisting}
system = dimos.AmberForceField("config.prmtop")
integrator = dimos.LangevinDynamics(dt, T, gamma, system)
simulation = dimos.MDSimulation(system, integrator, positions, T)
simulation.step(num_steps)
\end{lstlisting}
\end{minipage}
\end{center}
Here, a molecular system configuration from a common input format (\verb|config.prmtop|) is loaded and an integrator object to perform Langevin dynamics is initialized, which combined form the simulation; we then run the simulation for a given number of steps.
All other required functions are hidden to the user, such as the handling of boundary conditions or neighborlists, but they are still easily accessible for inspection or extension.
Due to the modular setup of the framework and since we are relying only on two external dependencies (\verb|PyTorch|\cite{NEURIPS2019_9015} to implement the simulation and \verb|parmed|\cite{shirts2017lessons} to read configuration files), it is straightforward to develop novel approaches, which we believe is a strength of our package. 
\par
In the remainder of this section, we compactly present the simulation components implemented in DIMOS, and discuss their use and \verb|Python| interface.
For all components we implement, we aim to reproduce the simulation behavior of OpenMM\cite{eastman2023openmm} and have checked, whenever possible, consistency with regard to their implementation both for static and dynamic properties (corresponding test scripts are part of the published code base).
Stochastic algorithms, such as the Langevin integrator, are implemented in a reproducible way by passing the random number generator object to every function relying on it.
\par
User-facing examples and the scripts to generate the data used in the manuscript are provided with the codebase, and the docomentation includes a guide for integrating custom MLIP models into DIMOS' PyTorch interface.

\subsection{Classical Force Fields}
\label{sec:FF}
In classical molecular mechanics modelling (MM) based on classical force fields, one distinguishes between non-bonded and covalently-bonded atoms, see Fig.~\ref{fig:schematic}(d), where the former interactions are commonly composed of pair-wise potentials of van der Waals type and electrostatic Coulomb interactions.
The interaction between bonded atoms, in addition to a pair-wise bond potential between atoms, is modeled by introducing higher-order interactions.
Some force-field parameterization additionally include corrections which are based on tabular definitions of interaction energies between four atoms, called energy correction maps (CMAP).\cite{brooks1983charmm}
\par
We implement the energy components needed to reproduce the most common parameterizations of AMBER\cite{weiner1981amber} class force fields, and some additional components often used in CHARMM type parameterizations.\cite{brooks1983charmm}

\subsubsection{Bonded Interactions}
The interaction between two atoms $i$ and $j$ that are covalently bonded is modeled by a simple harmonic pair-wise bond potential
\begin{equation}
    U_{\mathrm{bonded}}(i,j) = k_{ij}(r_{ij} - r_{ij}^0)^2,
\end{equation}
where $r_{ij} = | \vec{r}_{ij}|$ is the Euclidean distance between the atoms, $k_{ij}$ models the interaction strength and $r_{ij}^0$ is the equilibrium distance.
Usually, $k_{ij}$ and $r_{ij}^0$ depend on the atom types involved in the bond and not explicitly on the particular atom, but we still use this more general notation for simplicity.
\par
For pairs of three atoms $i$, $j$, and $l$, linked sequentially by covalent bonds, one likewise defines a harmonic potential, which now depends on the angle between the atoms 
\begin{equation}
    U_{\mathrm{angle}}(i,j,l) = k_{ijl}(\theta_{ijl} - \theta_{ijl}^0)^2,
\end{equation}
where using the unit vector $\hat{r}_{ij} = \vec{r}_{ij}/|\vec{r}_{ij}|$ one defines $\theta_{ijl} = \arccos (\hat{r}_{ij} \cdot \hat{r}_{lj}) $ as the angle between the three atoms, and the other parameters are defined in analogy to the harmonic bonded potential.
For some force-fields, the three-body interactions are described by the Urey-Bradley potential, which is a combination of the angle and bond potential.
\par
Interactions between four sequentially bonded atoms $i,j,l,m$ are described by a torsion potential
\begin{equation}
    U_{\mathrm{torsion}}(i,j,l,m) = \sum_n k_{ijlm}^n \left[ 1 + \cos (n \omega_{ijlm} - \omega_{ijlm}^0) \right],
\end{equation}
where $\omega_{ijlm}$ is the dihedral angle between the two planes spanned by $i,j,l$ and $j,l,m$ and $n$ specifies the multiplicity.
The other parameters are defined in analogy to the previous two contributions.
For the torsion parameters, one distinguishes between proper and improper dihedral angles, differing whether the involved atoms are in a sequential linear chain or not.
Alternatively, some force fields simply model the improper dihedrals harmonically.
\par
Implementing new energy contributions similar to the ones above is straightforward in DIMOS, demonstrating a key advantage of our package.
The only requirement is that any new class accepts the positions of the system as input and returns the final energy, see our example implementation of \verb|calc_energy| for the harmonic bond potential:
\begin{center}
\begin{minipage}{\mywidth}
\begin{lstlisting}
def calc_energy(self, pos):
    distances = get_distances_edge_list(
    pos, self.bond_list, self.periodic, self.box)
    return torch.sum(self.bond_parameters[:, 0] * 
    (distances - self.bond_parameters[:, 1])**2)
\end{lstlisting}
\end{minipage}
\end{center}
Here, \verb|get_distances_edge_list| takes care of returning the distances respecting periodic boundaries of the atoms defined in the \verb|self.bond_list|.
The parameters of the potential are stored in \verb|self.bond_parameters|, which is handed over at initialization of the class.

\subsubsection{Nonbonded Interactions}
In many simulations, the most time consuming part of the calculations is related to the non-bonded interactions.
\emph{A priori}, all atoms interact with each other, leading to $O(N^2)$ complexity for the computation of energy or forces.
The introduction of a cut-off allows one to use specialized data structures to store pairs of atoms that are within the interaction radius of each other, so-called neighborlists (typically called edge lists in the ML literature).
Then, the complexity of the energy or force computation becomes $O(N)$ or $O(N \log N)$, depending on the selected approximation scheme.
In the following, we will discuss the components of neighborlists and their implementation in \mbox{DIMOS}.

\paragraph{Neighborlists:}
In DIMOS, we employ a combination of Verlet neighborlist\cite{verlet1967computer} and cell lists.\cite{tildesleybook}
Neighborlists are data structures in which one simply stores all pairs of atoms that are within a distance $r_{\mathrm{nl}} = r_c + r_g$, where $r_c$ is the cutoff distance of the potential, and $r_g$ is the so-called ghost radius\cite{verlet1967computer} (set to $0.25r_c$ by default in DIMOS).
This implies that the list contains more atom pairs than strictly needed for the computation of the interactions and one wastes some computation when calculating the distance $r$ for all pairs with $r_{\mathrm{nl}}>r>r_c$.
However, this is more than compensated by being able to reuse the neighborlist more than once during the simulation.
This is because each atom only moves a relatively small distance at each iteration.
Due to the introduction of the ghost radius $r_g$, the pair list only looses its validity whenever a single atom has moved a distance $d$ greater than half the ghost radius $d>r_g/2$ since construction of the list.
Only then, the neighborlist needs to be reconstructed.
This reduces the complexity to $O(N)$ per synchronous update of all atoms whenever the neighborlist is valid, in contrast to an $O(N^2)$ complexity otherwise.
\par
However, a naive re-construction of the neighborlist still has $O(N^2)$ complexity, ultimately not improving the complexity of the simulation.
This is because the distance between all pairs of atoms needs to be calculated to decide which ones are smaller than $r_{\mathrm{nl}}$.
An alluring solution to this problem is the partitioning of the simulation box into cells based on the spatial coordinates of the atoms.
Each cell is chosen to be of at least $r_{\mathrm{nl}}$ length, and each atom is assigned to one of the cells.
Then, to find all atom pairs within $r_{\mathrm{nl}}$, one simply needs to check the neighboring cells (including off-diagonal neighbors) and store all combinations of atoms within the neighborlist.
The overall complexity of the algorithm is $O(N)$.
\par
By default, the end-user of DIMOS does not need to explicitly consider the use of neighborlists, since these are automatically been utilized and kept up-to-date.
For some applications, as for example training an MLIP on some pre-existing data, one can obtain the explicit list of atom pairs as shown in the following code snippet:
\begin{center}
\begin{minipage}{\mywidth}
\begin{lstlisting}
nb_handling = dimos.NeighborHandling(periodic, num_atoms, cutoff, box, exclusions)
nblist = nb_handling.get_neighborlist(pos)
\end{lstlisting}
\end{minipage}
\end{center}
\par
In case the boundary conditions are not periodic, the system is, for the sake of neighborlist computation, embedded into a non-periodic box.
For these new (virtual) boxes, the usual routines are called without wrapping around the coordinates of the cells as would be otherwise necessary with periodic boundaries.

\paragraph{Van der Waals interactions:}
One of the most used functions to model van der Waals interactions is the $12$--$6$ Lennard Jones potential
\begin{equation}
U_{\mathrm{LJ}}(r_{ij}) = 4 \epsilon_{ij} \left( \left( \frac{\sigma_{ij}}{r_{ij}} \right)^{12} - \left( \frac{\sigma_{ij}}{r_{ij}} \right)^6 \right).
\label{eq:lj}
\end{equation}
The parameters $\epsilon_{ij}$ and $\sigma_{ij}$ are determined from atom specific parameters $\epsilon_i$ and $\sigma_i$ by the Lorentz-Berthelot combining rules\cite{lorentz1881ueber,berthelot1898melange} as
\begin{equation}
\epsilon_{ij} = \sqrt{\epsilon_i\epsilon_j},\,\,\,\,\sigma_{ij} = \frac{\sigma_i + \sigma_j}{2}.
\end{equation}
Alternatively, also fully geometric combining can be specified.
\par
To avoid non-smooth behavior at the cutoff $r_c$, one can adapt different schemes.
We here follow the approach of OpenMM\cite{eastman2023openmm} and apply the switching function
\begin{equation}
    S(x) = 1 - 6x^5 + 15x^4-10x^3
\end{equation}
to Eq.~\eqref{eq:lj}, where $x=(r_{ij}-r_s)/(r_c-r_s)$ and $r_s<r_c$ is the switching distance.
For simulations at constant temperature and pressure, we have also implemented long-range dispersion corrections that approximate the contributions beyond $r_c$.\cite{shirts2007accurate}
\paragraph{Coulomb interactions:}
The coulomb potential used to model electrostatic interactions is given by
\begin{equation}
U_{\mathrm{el}}(r_{ij}) = \frac{1}{4\pi\upepsilon_0} \frac{q_iq_j}{r_{ij}},
\label{eq:coulomb}
\end{equation}
where $\upepsilon_0$ is the vacuum permittivity and $q_i$ is the (partial) charge of atom $i$.
Similar to the Lennard-Jones case, it is possible to introduce a direct cutoff in real space.
This approach is referred to as the reaction field approximation\cite{tironi1995generalized} (RFA) which modifies Eq.~\eqref{eq:coulomb} to read
\begin{equation}
    U_{\mathrm{RFA}}(r_{ij}) = \frac{1}{4\pi\upepsilon_0} q_i q_j \left( r_{ij}^{-1} + k_{\mathrm{RFA}} r_{ij}^2 - c_{\mathrm{RFA}} \right)
    \label{eq:RFA}
\end{equation}
with $k_{\mathrm{RFA}} = r_c^{-3} (\upepsilon_s-1) / (2\upepsilon + 1)$ and $c_{\mathrm{RFA}} = 3 \upepsilon_s r_c^{-1}  / (2 \upepsilon +1)$, where $\upepsilon_s$ is the dielectric constant of the solvent.
This approach thus approximates the influence of the atoms outside the cutoff in a mean-field manner, and may consequently introduce uncontrolled effects.
\par
A less heuristic method to introduce a cutoff in electrostatic modelling is Ewald summation,\cite{toukmaji1996ewald} for which one writes
\begin{equation}
    U_{\mathrm{Ewald}}(r_{ij}) = U_{\mathrm{dir.}}(r_{ij}) + U_{\mathrm{recip.}}(r_{ij}) + U_{\mathrm{self}}
\end{equation}
with
\begin{eqnarray}
    &U&_{\mathrm{dir.}} = \frac12 \sum_{i,j} \sum_{\vec{n} \in \bf n} q_i q_j \frac{\erfc (\alpha |\vec{r} + \vec{n}|)}{|\vec{r} + \vec{n}|}\\
    &U&_{\mathrm{recip.}} = \frac{1}{2\pi V} \sum_{i,j} \sum_{\vec{k} \in {\bf k} \neq 0} \frac{\exp(-(\pi \vec{k}/\alpha)^2 + 2\pi \vec{k}\cdot\vec{r}_{ij})}{\vec{k}^2}\\
    &U&_{\mathrm{self}} = -\frac{\alpha}{\sqrt{\pi}} \sum_i q_i^2.
\end{eqnarray}
The potential is thus split into a real space contribution $U_{\mathrm{dir.}}$, a reciprocal space contribution $U_{\mathrm{recip.}}$, and the self energy $U_{\mathrm{self}}$.
The direct and reciprocal space sums run over all periodic copies of the system, respectively, all reciprocal vectors.
$V$ is the volume of the simulation box, and $\alpha$ controls the shift between $U_{\mathrm{recip.}}$ and $U_{\mathrm{self}}$.
\par
By construction, the sums of the individual energy contributions in real and reciprocal space converge significantly faster than directly evaluating the sum.
Hence, it becomes possible to introduce truncations of both sums, in practice introducing a cut-off $r_c$ in real, respectively, $k_c$ in reciprocal space.
Compared to the naive sum, a much smaller cutoff for both sums can be used to reach a target accuracy, significantly reducing the number of necessary computations.
\par
For an ideal choice of cutoff of the real and reciprocal summation as well as crossover parameter $\alpha$, it can be shown that this reduces the computational effort from $O(N^2)$ to $O(N^{3/2})$ per step.
Note that for any choice of $\alpha$ above formulas are correct; hence, the choice of $\alpha$ only influences which cutoff $r_c$ and $k_c$ can be chosen for a desired accuracy.
In practice, $\alpha$ is by default heuristically determined from the prescribed tolerance $\delta$ on the energy approximation and a fixed cutoff $r_c$ (typically the one used for the van der Waals interactions) as\cite{eastman2023openmm}
\begin{equation}
\alpha = \sqrt{-\log 2\delta}/r_c.
\label{eq:alpha}
\end{equation}
Then, one uses a root-finding algorithm to determine $k_c$ via the empirically found relation 
\begin{equation}
\delta = \frac{k_c \sqrt{d \alpha}}{20} \exp(-(\pi k_c/d\alpha)^2).
\end{equation}
\par
To further speed-up computation, it is possible to carry out the reciprocal sum using fast Fourier transform (FFT) by distributing the charges onto a lattice via spline interpolation.
This significantly speeds up the calculation of $U_{\mathrm{recip.}}$, and the overall complexity reduces to $O(N \log N)$.
Following OpenMM, we implement smooth particle-mesh Ewald (PME).\cite{essmann1995smooth,eastman2023openmm}
For this, we also implement B-splines\cite{bartels1995introduction} based on the recursive definition of the splines, resulting in custom forward and backward implementation within \verb|PyTorch|.
The number of nodes in the mesh is chosen heuristically as $n_{\mathrm{mesh}} = 2\alpha d/3 \delta^{1/5}$, where $\alpha$ is again determined from Eq.~(\ref{eq:alpha}).

\paragraph{Exclusions and Exceptions:}
Atoms that are directly bonded or indirectly bonded by one additional covalent bond to each other are excluded from the non-bonded energy terms.
In addition, atoms connected via three bonds (such as in torsion contributions) are often treated differently.
The most common way to handle these $1$-$4$ exceptions is to scale the strength of the non-bonded interactions by some common factor $<1$.
Alternatively, some parameterizations of force-fields scale each exclusion by an individual factor.
In DIMOS, exclusions and exceptions are handled at the level of neighborlists, where excluded interaction pairs are filtered out.
Since all possible $1$-$4$ exceptions are known at simulation setup, they are handled explicitly by the non-bonded interaction method. 

\subsection{Machine Learning Based Descriptions of Intermolecular Interactions}
\label{sec:ML_enhanced}
MLIPs do typically not introduce an explicit distinction between non-bonded and bonded atoms, and instead often build higher-order interactions between all atoms within a cutoff distance using various approaches, see Ref.~\onlinecite{duval2023hitchhiker} and references therein.
Usually, these approaches include physical features in some capacity, for example a large body of work only considers features that are invariant to rotations and translations of the system and uses rather standard ML architectures with learnable parameters.\cite{schutt2017schnet}
Another line of work explicitly constructs the architectures such that their internal features or their output transform equivariantly to the input, i.e., for example rotating the system also leads to a predictable rotation of the output.\cite{Batatia2022mace,neumann2024orbfastscalableneural,zaverkin2024higher}
\par
We implement the interface to two current state-of-the-art MLIP models: MACE\cite{Batatia2022mace} and ORB.\cite{neumann2024orbfastscalableneural}
We want, however, to stress that DIMOS makes no explicit assumption on the model and is rather agnostic to the details, as long as the models are implemented in \verb|PyTorch|.
In Fig.~\ref{fig:schematic}(b) we schematically demonstrate the basic setup of message-passing graph neural network MLIPs, in which many-body interaction terms are built using successive message passing iterations.
However, since we do not implement any novel MLIP architecture in DIMOS, we refer to the literature on details as well as advantages/disadvantages of the respective approaches.\cite{ko2023recent,duval2023hitchhiker}
\par
Using DIMOS, the best performance is achieved if the model expects a combination of the neighborlist, distance vectors or distances, and/or shifts, since they are calculated anyway in DIMOS and recalculating (some of them) within the model leads to wasted calculations.
By having the full simulation framework implemented in \verb|PyTorch|, it is possible to compile parts or the whole simulation.
This also extends to compiling MLIPs using \verb|torch.compile|, which is currently not possible when using LAMMPS\cite{LAMMPS} or OpenMM\cite{eastman2023openmm} to interface the MLIP.
Instead, these simulation engines require the use of \verb|TorchScript|, forcing users to rewrite model code to conform to a restricted subset of Python in order to convert the model into a static graph.
Albeit this does not necessarily impose a performance penalty, it makes the code more rigid and often harder to debug, due to loosing many of Python's dynamic features.
\par
Setting up a system modeled by a (foundational) MACE model can be achieved in a single line:
\begin{center}
\begin{minipage}{\mywidth}
\begin{lstlisting}
ml_system = dimos.MaceFoundationalSystem(masses, atomic_numbers)
\end{lstlisting}
\end{minipage}
\end{center}
In addition, there are some optional parameters, which, however, are not necessary in general.
This highlights the simplicity of using DIMOS to run simulations using an MLIP, compared to other established approaches, often requiring a larger number of intermediate steps.
For example, using the model in LAMMPS\cite{LAMMPS} requires implementing it using \verb|TorchScript| in the first place (as discussed above) and compiling LAMMPS from scratch using either the ML-IAP extension or the specific branch maintained by the MACE developers, then exporting the model in a specific way, generating a corresponding input file, and transforming the coordinate file into a usable format.
\par
MACE is currently one of the most prominent MLIPs in the literature,\cite{Batatia2022mace} with diverse applications in different domains.
We specifically employ the MACE material project foundation model (\verb|mace-mp|),\cite{batatia2023foundation} which is available in the \verb|small|, \verb|medium|, and \verb|large| variant depending on the chosen parameterization.
In DIMOS, we support the acceleration of MACE using NVIDIA cuEquivariance,\footnote{See https://github.com/NVIDIA/cuEquivariance for more details.} and activate it in all our benchmarks.

\subsection{Machine Learning/Molecular Mechanics}
An alternative to describing the system fully using an MLIP is to employ a hybrid approach, where part of the system is described using an MLIP, and the remaining part by a classical force field.
This is often called ML/MM, standing for machine learning/molecular mechanics, a term derived from QM/MM used to describe the combination of quantum mechanical \emph{ab initio} calculations and molecular modelling.
While the field of QM/MM is quite advanced, in particular in choosing a proper embedding approach for the electrostatic interactions,\cite{lin2007qm} it is significantly less explored for the ML/MM approach.
The current approaches are based on a mechanistic embedding, and as such the electrostatic interactions are treated on the MM level\cite{galvelis2023nnp} or ignored.
To define an ML/MM system, one defines a \verb|ml_system| which is modeled with an MLIP and a \verb|mm_sytem| described by a classical force field.
In addition, one needs define \verb|ml_atoms|, a list containing the atom indices modeled by the \verb|ml_system|, which then results in the new combined system.
More details are presented in Section~\ref{sec:benchmark_mlips}, including example code on how to achieve this is (see Fig.~\ref{fig:benchmark_protein}(a)).

\subsection{Molecular Dynamics and Energy Minimization}
\label{sec:integrators}
With the system description in place, we now present the implemented methods that update the position of the atoms, either to perform molecular dynamics simulations or minimize the potential energy.

\subsubsection{Molecular Dynamics}
\label{sec:MD}
The basis of MD is numerically solving Newtons equation of motion given by $\frac{\partial E}{\partial \vec{r}} = -\vec{F} = -m\vec{a}$.
To do so, we implement the ubiquitous methods by Verlet\cite{verlet1967computer} in its velocity variant,\cite{swope1982computer} which numerically integrates Newtons equation in time and hence conserve the total energy of the shadow Hamiltonian.
\par
An MD simulation utilizing the velocity Verlet integrator consists of the following steps:
\begin{center}
\begin{minipage}{\mywidth}
\begin{lstlisting}[backgroundcolor = \color{nearlywhite}, mathescape]
1. Update velocities: $\vec{v}_i(t+\frac12 \Delta t) = \vec{v}_i(t) + \frac12 \vec{a}_i(t)\Delta t$.
2. Update positions: $\vec{x}_i(t+\Delta t) = \vec{x}_i(t) + \vec{v}_i(t + \frac12 \Delta t)\Delta t$.
3. Apply position constraints.
4. Calculate $\vec{a}_i(t+\Delta t)$.
5. Perform second velocity update: $\vec{v}_i(t + \Delta t) = \vec{v}_i(t + \frac12 \Delta t) + \frac12 \vec{a}_i(t + \Delta t)\Delta t$.
6. Apply velocity constraints.
7. Apply thermostat.
8. Measure properties.
9. Go to 1.
\end{lstlisting}
\end{minipage}
\end{center}
Here, $\vec{x}_i$ is the position of atom $i$, $\vec{v}_i$ and $\vec{a}_i$ are the velocity and acceleration.
Each step is understood to be performed for all atoms $i$.
Since the updates are performed synchronously, the update for different atoms can easily be performed in parallel.
In step \verb|3| and \verb|6| the respective constraint algorithms that fix the distances between atoms or angles to their target value is run.
\subsubsection{Constant Temperature Simulations}
The most direct way to control the temperature of a system, i.e., to sample the canonical ensemble, are thermostats, applied in step \verb|7|.
For an overview on different thermostatting strategies, see Ref.~\onlinecite{Hünenberger2005} and references therein.
We implement the Berendsen, Andersen, and Lowe-Andersen thermostat.\cite{koopman2006advantages}
These thermostats work by either rescaling the atoms velocities or by picking new velocities from the Maxwell-Boltzmann distribution for a fraction of atoms or pairs of atoms.\cite{koopman2006advantages}
\par
It is also possible to define equations that deviate from Newton's equation of motion, leading for example to the Langevin and Brownian dynamics equation.\cite{Hünenberger2005}
These integrators require adapted integration schemes, which we have also implemented in DIMOS.\cite{zhang2019unified}
\par
Another way to control the temperature is to introduce additional degrees of freedom that act as coupling to an external heatbath.
This also modifies the integrator, where now the additional degree of freedoms are explicitly integrated using specific splitting schemes.
We implement the Nos\'{e}--Hoover and Nos\'{e}--Hoover chain dynamics,\cite{Hünenberger2005} which introduce one or more additional harmonic `particles' that couple the system's temperature to the target temperature.
\par
Practically, in DIMOS, one defines all different type of thermostats or ways to perform the dynamics as separate integrators.
Below, as example, we present the code necessary to define the Langevin and Nos\'{e}--Hoover chain dynamics, as well as dynamics using the velocity Verlet integrator in conjunction with the Andersen thermostat.
\begin{center}
\begin{minipage}{\mywidth}
\begin{lstlisting}
integrator = dimos.LangevinDynamics(dt, T, gamma, system)
integrator = dimos.NoseHooverChainDynamics(dt, T, freq, system)
integrator = dimos.AndersenDynamics(dt, T, freq, system)
\end{lstlisting}
\end{minipage}
\end{center}
\subsubsection{Constant Temperature and Pressure Simulations}
To reproduce many experimental conditions correctly, it is necessary to simulate the system at constant pressure.
We implement an isotropic and anisotropic Monte Carlo barostat,\cite{aaqvist2004molecular} which works by proposing new box sizes and accepting the new state with the usual Metropolis Monte Carlo criterion, guaranteeing that the stationary distribution of samples is the target distribution.
In particular, one proposes new box sizes for the simulation cell, and then repositions the coordinates of the atoms within this box to correspond to the changed size.
A high acceptance rate is achieved by displacing the center of each molecule instead of treating each atom individually, as this way one avoids distortions within each molecule.
Alternatively, in cases where molecules are not as clearly defined such as for MLIPs, it is also possible to simply displace every atom independently.
\par
In DIMOS, the desired barostat is defined, and then passed together with the \verb|system| and \verb|integrator| to initialize the simulation object.
\begin{center}
\begin{minipage}{\mywidth}
\begin{lstlisting}
barostat = dimos.MCBarostatIsotropic(box, target_pressure, freq)
simulation = dimos.MDSimulation(system, integrator, positions, barostat, T)
\end{lstlisting}
\end{minipage}
\end{center}

\begin{figure*}
  \includegraphics[width=0.32\textwidth]{{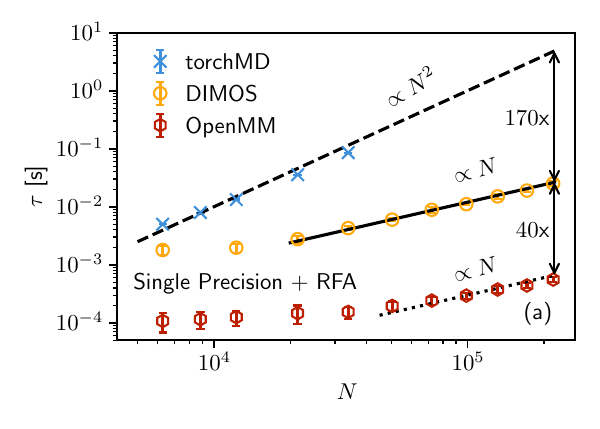}}
  \includegraphics[width=0.32\textwidth]{{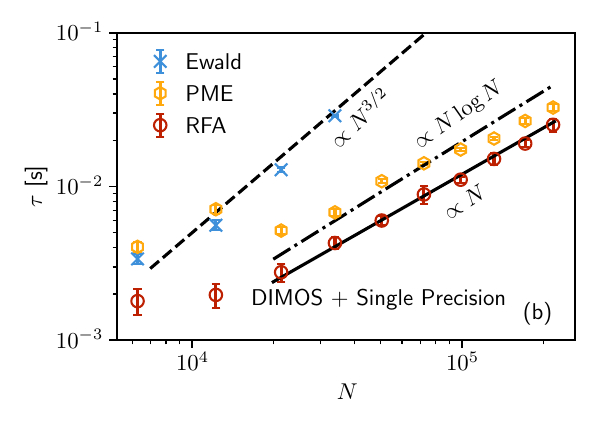}}
  \includegraphics[width=0.32\textwidth]{{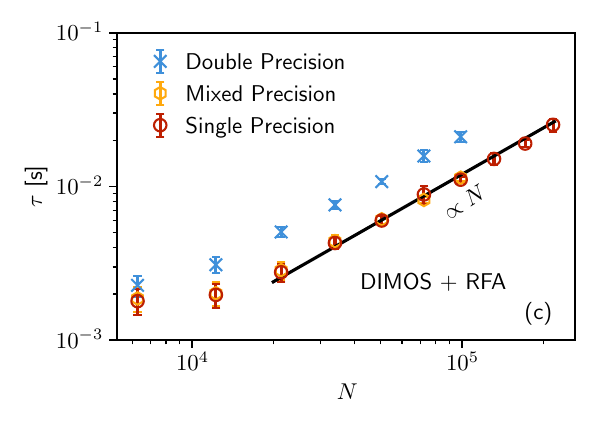}}\\
  \includegraphics[width=0.32\textwidth]{{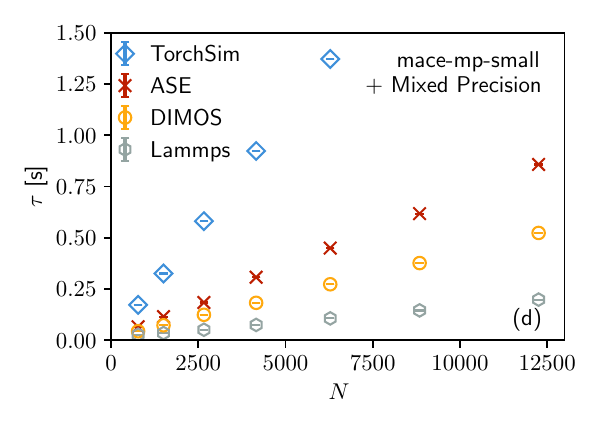}}
  \includegraphics[width=0.32\textwidth]{{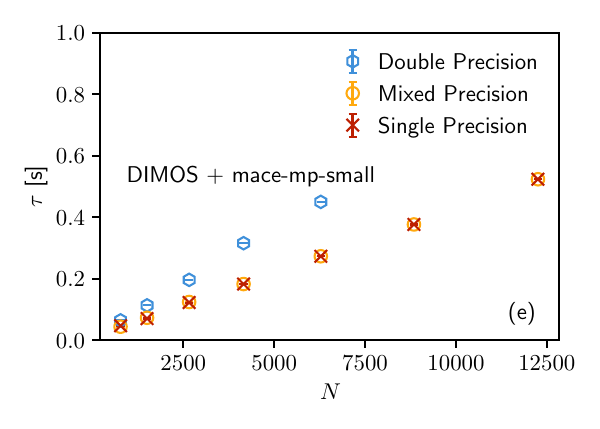}}
  \includegraphics[width=0.32\textwidth]{{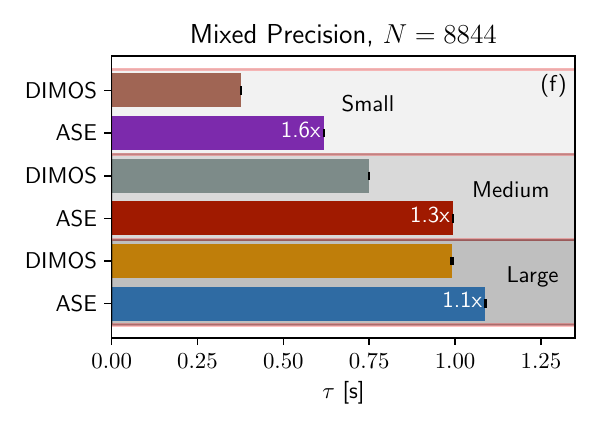}}
  \caption{Runtime $\tau$ in seconds per MD step for a water box with constant density as a function of number of atoms $N$. (a) Shows the runtimes for torchMD, DIMOS, and openMM using single precision numerics and RFA to model the electrostatic interactions. (b) Demonstrates the influence of different approaches to model the electrostatic interactions: Runtimes are presented for RFA, Ewald, and PME for DIMOS using single precision numerical accuracy. In (c), we investigate the effect of the chosen computational precision on the runtime, modelling electrostatic interactions via RFA in DIMOS.
  In all cases, the lines indicate the theoretically expected asymptotic computational complexity.
  (d)-(f) Show the performance of running the MACE foundation model, where in (d) we compare the runtimes of DIMOS to TorchSim, ASE, and LAMMPS using the small MACE model and mixed precision (apart from TorchSim, which uses single precision) for different $N$, and (e) investigates the influence of choosing different numerical precision within DIMOS.
  (f) Presents the runtimes for the biggest system that can be simulated using the large MACE model ($N=8844$) and compares the resulting $\tau$ for DIMOS and ASE.}
  \label{fig:benchmark}
\end{figure*}
\subsubsection{Bond and Angle Constraints}
\label{sec:methods_constraints}
To enable a larger time step $\Delta t$ in integration, it is common\cite{perez2015constraint} to restrict the highest-frequency motions of the system, and hence to fix the involved relative positions of the atoms to their equilibrium value.
This effectively allows increasing $\Delta t$ such that the second-fastest motion is sampled with enough intermediate observations between oscillations.
\par
We implement two constraint algorithms, SETTLE\cite{miyamoto1992settle} and the constant constraint matrix approximation (CCMA) approach.\cite{eastman2010constant}
The former method constraints both positions and velocities by finding analytical expression for three body systems with three constraints, typical for parameterization of water molecules, such as TIP3P.\cite{jorgensen1983comparison}
CCMA, on the other hand, is an iterative method that is more generally applicable for arbitrary constraint topologies.
It is an iterative Lagrange multiplier-based method, in which the (inverse) of the Jacobian matrix is approximated to be constant by utilizing equilibrium expectation values.
\par
To use constraint algorithms, DIMOS provides a keyword based option for the most common type of constraints when generating the \verb|system|, i.e., either constraining all bonded interactions containing at least one hydrogen or additionally constraining all angles where at least one of the atoms is a hydrogen.
\begin{center}
\begin{minipage}{\mywidth}
\begin{lstlisting}
system = dimos.AmberForceField(filename, constraint_option="h_bonds"|"h_angles")
\end{lstlisting}
\end{minipage}
\end{center}
If any of the options is activated, water molecules are always constrained, following the convention of OpenMM.
Whether to use SETTLE or CCMA for the list of constrained atoms is automatically determined, and does not need to be specified by the user.
\par
If a more fine-grained control is desired, it is possible to explicitly set the \verb|constraint_handler| object of the \verb|integrator|:
\begin{center}
\begin{minipage}{\mywidth}
\begin{lstlisting}
integrator.constraint_handler = dimos.ConstraintHandling(
            constraints, constr_eq_val,
            masses, num_atoms, angle_list,
            angle_eq_parameters, tolerance)
\end{lstlisting}
\end{minipage}
\end{center}

\subsubsection{Energy minimization}
We implement energy minimization by utilizing the optimizer framework available in \verb|torch.optim|, usually employed for learning network parameters via back-propagation.
This allows for easy access to methods like gradient descent or variants thereof including momentum terms, such as commonly used in ML.
\par
By default, we utilize limited-memory BFGS,\cite{liu1989limited} which is also standard in OpenMM.\cite{eastman2023openmm}
\par
Having already defined a simulation object, the energy can be minimized simply by calling
\begin{center}
\begin{minipage}{\mywidth}
\begin{lstlisting}
simulation.minimize_energy(steps, optimizer="LBFGS")
# If a custom optimizer is used:
def optim(parameters): 
    return torch.optim.Adam(parameters, lr=lr)
simulation.minimize_energy(steps, optimizer=optim)
\end{lstlisting}
\end{minipage}
\end{center}
This allows also the testing of some advanced optimizers developed in the ML literature to be evaluated on molecular systems.

\section{Performance Benchmarks}
\label{sec:benchmark}
In the following section, we present the resulting runtimes of DIMOS for simulations of water boxes to probe the influence of system size and simulation/modelling choices such as the method to calculate long-range interactions or different schemes for the numerical datatypes for classical force fields and MLIPs.
Then we turn towards the simulation of proteins in explicit solvent, investigating the influence of constraints on the simulation speed and presenting ML/MM as hybrid solution.
Finally, we demonstrate how the end-to-end differentiability of the code base allows us to differentiate through the whole simulation to tune parameters of the simulation and system in the framework of HMC.\cite{duane1987hybrid, neal2011}
\par
All simulation results for which we quote runtimes were performed using an NVIDIA RTX 6000 Ada GPU with 48GB memory on a server equipped with an Intel Xeon Gold 5412U CPU with 24 cores and 256GB memory.

\begin{figure*}
    \begin{minipage}{0.85\textwidth}
    \begin{lstlisting}
ml_atoms = [...] # Indices of atoms that are being modeled by ML system.
mm_system = dimos.GromacsForceField("config.top", "config.gro", excluded_bonded_atoms=ml_atoms)
ml_system = mm_system.as_mace_system("mace-mp-large", atom_range=[0,len(ml_atoms)])
mlmm_system = dimos.mlmm.MLMMsystem(ml_atoms, ml_system, mm_system)
    \end{lstlisting}
    \raggedright
    (a) Example code showing how to set up a simulation of an ML and ML/MM modeled system starting from an MM description.
    \end{minipage}
    \includegraphics[width=0.32\textwidth]{{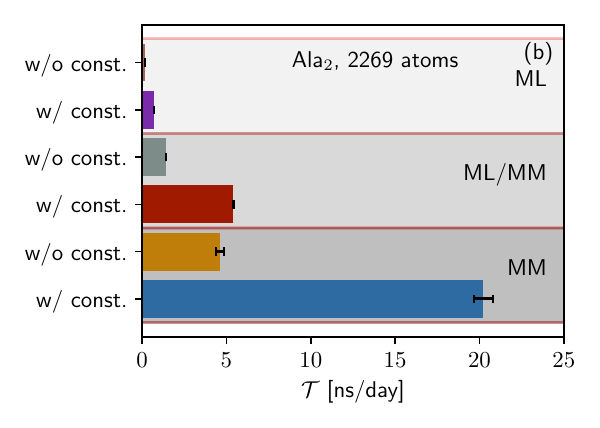}}
    \includegraphics[width=0.32\textwidth]{{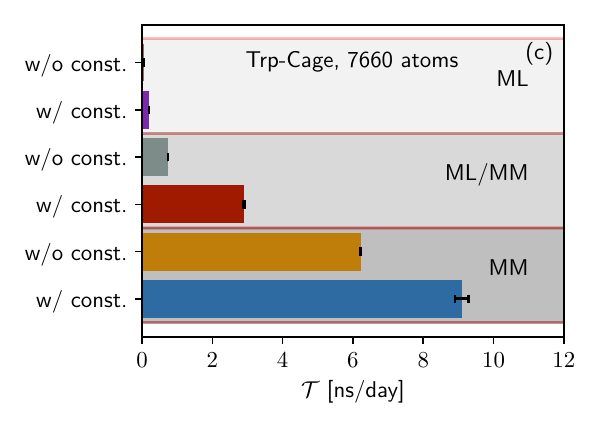}}
    \includegraphics[width=0.32\textwidth]{{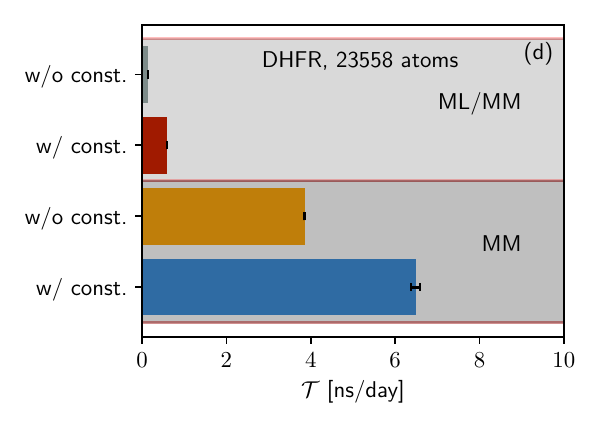}}\\
    \cprotect\caption{(a) Demonstration on how to set up both classically and machine-learning enhanced systems in DIMOS. \verb|ml_atoms| specifies which atoms should be modeled by the MLIP system, and also can be empty in case one does not want to model any systems using ML. (b-d) Achievable simulation time per day $\mathcal{T}$ with and without constrained motion of the hydrogens for each of the three cases of a system modeled purely by ML, the hybrid approach using ML/MM, and a purely MM based modelling. In all cases, the MLIP model was chosen to be \verb|mace-mp-large|. In (b), we present the data for alanine-dipeptide (Ala$_2$), (c) shows Trp-Cage, and (d) shows dihydrofolate reductase (DHFR). All systems are modeled using explicit water modeled by the TIP3P parameters.\cite{jorgensen1983comparison} In (d), the biggest system, it is not possible to model purely with ML, as for this system size the MLIP model required more than the available GPU memory.}
    \label{fig:benchmark_protein}
\end{figure*}

\subsection{System-size Scalability}
\label{sec:benchmark_water}
We investigate the runtimes per simulation step averaged over several simulation steps for water boxes of fixed density with varying number of atoms $N$.
In Figs.~\ref{fig:benchmark}(a)-(c), we present the average runtime per MD step $\tau$ as a function of number of atoms $N$ in an unconstrained simulation of water under Langevin dynamics at $300$~K with timestep $\Delta t=0.5$~fs, friction coefficient $\gamma=0.1$~fs$^{-1}$, and cutoff $r_c=10$ \r{A}.
The water is modeled by the TIP3P water model.\cite{jorgensen1983comparison}
\par
In (a) we compare $\tau$ obtained in DIMOS to torchMD\cite{doerr2021torchmd} and openMM\cite{eastman2023openmm} for simulations running in single floating point precision and electrostatic interactions modeled by the reaction field approximation (the RFA is the only setup for electrostatics supported by torchMD).
We observe linear scaling of the runtime per step $t$ in DIMOS and openMM, whereas torchMD displays quadratic scaling as a function of number of atoms $N$.
For the largest system size tested ($\approx 300,000$ atoms), DIMOS as a result achieves a $170\times$ speed-up over torchMD, but is factor $40\times$ slower than openMM. 
For torchMD, the runtime of the largest system size is extrapolated from the expected quadratic complexity, while, as expected, DIMOS and openMM have linear scaling due to utilizing neighborlists and cell lists.\footnote{For a fair comparison, we check the validity of the neighborlist after every integration step, avoiding any potential artifacts leading to inaccuracies in the force/energy evaluation.}
The slowdown compared to openMM is epected, since DIMOS is implemented on a much higher abstraction level prohibiting many optimizations.
This aligns with our envisioned role of DIMOS as a complementary platform for rapid prototyping and gradient-based workflows rather than a direct replacement for production-optimized codes
\par
Panel~(b) demonstrates the influence of different methods how to model electrostatics within DIMOS.
We use single precision numerics.
Clearly the RFA is the fastest option due to the simple mean-field treatment of charges outside the cutoff.
Somewhat slower, but still competitive is PME,\cite{essmann1995smooth} for which one expects $O(N \log N)$ complexity, which roughly agrees with our data.
In contrast, using direct Ewald summation\cite{toukmaji1996ewald} with theoretically ideal complexity $O(N^{3/2})$ is significantly slower already for medium-sized systems, and can only provide a minor performance improvement over PME for small systems.
In addition, direct Ewald summation requires significantly more GPU memory and can thus not be used for large systems.
\par
The influence of the chosen numerical accuracy is presented in Fig.~\ref{fig:benchmark}(c) with electrostatic interactions modeled by RFA, where we find that double precision as expected is overall slower on our GPU.
Very little performance penalty is observed using mixed precision, where the integration is performed in double precision, but the force/energy evaluation is performed in single precision.
However, this approach cannot be used for larger systems, as the memory requirements increase significantly.
If memory consumption is not a bottleneck, it is advisable to always consider using mixed precision due to the reduction of accumulation of errors in comparison to single precision, where such effects can be detrimental.
\par
In Figs.~\ref{fig:benchmark}(d)-(f) we investigate the behavior of DIMOS in conjunction with MLIPs by simulating the same water boxes as in (a)-(c) with the \verb|mace-mp| foundational model.\cite{Batatia2022mace,batatia2023foundation}
MACE is a state-of-the-art equivariant message-passing graph neural network that models interatomic interactions.
The simulation parameters remain largely the same, however, we now employ the standard interaction cutoff for message passing of $r_c=3$ \r{A} in MACE.
\par
In panel (d), we compare DIMOS to three different simulation engines:
\emph{i}) TorchSim,\cite{cohen2025torchsimefficientatomisticsimulation} a (differentiable) torch-based simulation package released after DIMOS focusing on batching simulations, \emph{ii}) ASE,\cite{ase-paper} one of the most popular (non-differentiable) approaches to run MLIP based simulation due to its simplicity, and \emph{iii}) LAMMPS,\cite{LAMMPS} using the ML-IAP interface as example of a compiled simulation engine, prioritizing simulation speed but, as discussed previously, incurring additional overhead in model setup.
Apart from TorchSim, for which we use single precision since mixed precision is not supported, all simulations are performed in mixed precision.
We observe that for the smallest foundation model \verb|mace-mp-small|, DIMOS positions itself as the fastest high-level simulation engine.
TorchSim suffers from slow simulation speed and large memory demands, the latter limiting the maximal system size we can simulate.\footnote{We follow the official TorchSim tutorial to set up the simulation.}
As a consequence, for the biggest system size that is simulated by TorchSim of $N=6282$, DIMOS is roughly factor $\approx 5$ faster than TorchSim.
DIMOS is also faster than ASE by factor $\approx 1.7$ for the biggest system size $N=12255$.
Compared to LAMMPS, DIMOS is roughly $2.5\times$ slower, but employing LAMMPS, as discussed, comes with constraints on the atomic descriptor and requires the model to be implemented using \verb|TorchScript|.
Consequently, for large-scale simulations, it may be advantageous to adopt a hybrid approach that leverages the respective strengths of both packages to maximize performance and flexibility.
\par
In (e), we present the influence of choosing different numerical precision, i.e., using single, mixed, or double precision.
There is virtually no penalty in using mixed precision, because the runtime and memory consumption is dominated by the ML model, so that all other factors effectively become negligible, at least for the currently achievable system sizes.
\par
Figure~\ref{fig:benchmark}(f) shows the influence of choosing different sizes of the \verb|mace-mp| model in comparison between DIMOS and ASE for the biggest system size ($N=8844$) that can be simulated using all ML model sizes.
We here focus on ASE, since TorchSim has higher memory requirements and slower simulation speed and LAMMPS is less directly comparable in user experience.
For the small model, the speed-up obtained by DIMOS is the biggest with factor $1.6\times$, i.e., nearly half of the time in ASE is spent in integrating the equation of motion and memory transfer.
For the bigger models, the difference between DIMOS and ASE become less pronounced due to the larger time spent in the actual ML model evaluation, but the speed-up still remains sizable.

\subsection{Constraints and Hybrid ML/MM for Protein Systems}
\label{sec:benchmark_mlips}
We simulate three protein benchmark systems with explicit solvent of varying size: \emph{i}) Alaninedipeptide (Ala$_2$) with $2269$ atoms, \emph{ii}) Trp-Cage with $7660$ atoms, and \emph{iii}) dihydrofolate reductase (DHFR) with $23558$ atoms.
The first two systems are standard test-systems to probe novel simulation methods, and the latter is a common testsystem from the joint AMBER/CHARMM benchmark.\footnote{See https://ambermd.org/GPUPerformance.php.}
Since we here also want to probe the effect of using constraint algorithms to increase the timestep on the achievable simulation time, we shift our metric to the simulation time achievable per wall-clock day $\mathcal{T}$, i.e., how many ns of simulation can be achieved per day.
In the simulations we use the Nos\'{e}--Hoover integrator with collision frequency of $0.05$~fs$^{-1}$ and $\Delta t = 0.5$~fs timestep for disabled constraints, and $\Delta t = 2.0$~fs timestep with constraints enabled.
\par
We are considering three scenarios: \emph{i}) The system purely modeled by \verb|mace-mp-large| (ML), \emph{ii}) a hybrid approach where the protein is modeled by the MACE model (ML/MM), but the water is parameterized by the classical TIP3P water model,\cite{jorgensen1983comparison} and \emph{iii}) a fully classical simulation using AMBER99SB\cite{weiner1981amber} force field and TIP3P water (MM).\cite{jorgensen1983comparison}
In the ML and ML/MM case, electrostatic contributions are ignored (DIMOS implements several options), whereas in the MM case we use PME.
As demonstrated in Fig.~\ref{fig:benchmark_protein}(a), it is straightforward to set up systems modeled by the three options in DIMOS.
The only option to be specified by the user is which MLIP to use, and which atoms should be modeled by it.
\par
In Fig.~\ref{fig:benchmark_protein}(b), we present the data for Ala$_2$ using the three approaches with and without constraints.
For the systems modeled by ML or ML/MM, we adopt the constraints as defined for the MM system.
The ML approach is clearly the slowest option with $\mathcal{T} \approx 0.18$~ns/day of simulation time without constraints.
This is in contrast to the fastest simulation time $\mathcal{T} \approx 20$~{ns/day} (more than $100\times$ faster than the ML approach), achieved by modelling the interaction using MM and with constraints activated.
As viable alternative, one may consider the hybrid approach, where with constraints we achieve $\mathcal{T} \approx 5.4$~ns/day.
\par
For the larger Trp-Cage system presented in (c), the differences become even more pronounced.
While a classical ML simulation without constraints achieves $\mathcal{T} \approx 0.05$~ns/day, the ML/MM approach without constraints is $14\times$ faster.
In the extreme case, employing constraints and modelling all interactions using MM, we observe a speed-up of $>175 \times$.
While modelling a part or all of the interactions using MM comes with an accuracy penalty, that this is easily possible to be implemented in DIMOS highlights the modularity of the framework with easy combination of various techniques tailored to the problem.
\par
Figure~\ref{fig:benchmark_protein}(d) shows the performance for DHFR, which due to memory constraints of MACE could not be simulated completely using the ML approach on our hardware.
Using ML/MM, we obtain runtimes in the range of $\approx 0.15$ to $\mathcal{T} \approx 0.6$~ns/day, and $\approx 3.8$ to $\approx 6.5$~ns/day for pure MM modelling.

\section{End-to-end Optimization of Hamiltonian Monte Carlo}
\label{sec:hmc}
\begin{figure*}
  \includegraphics[width=\textwidth]{{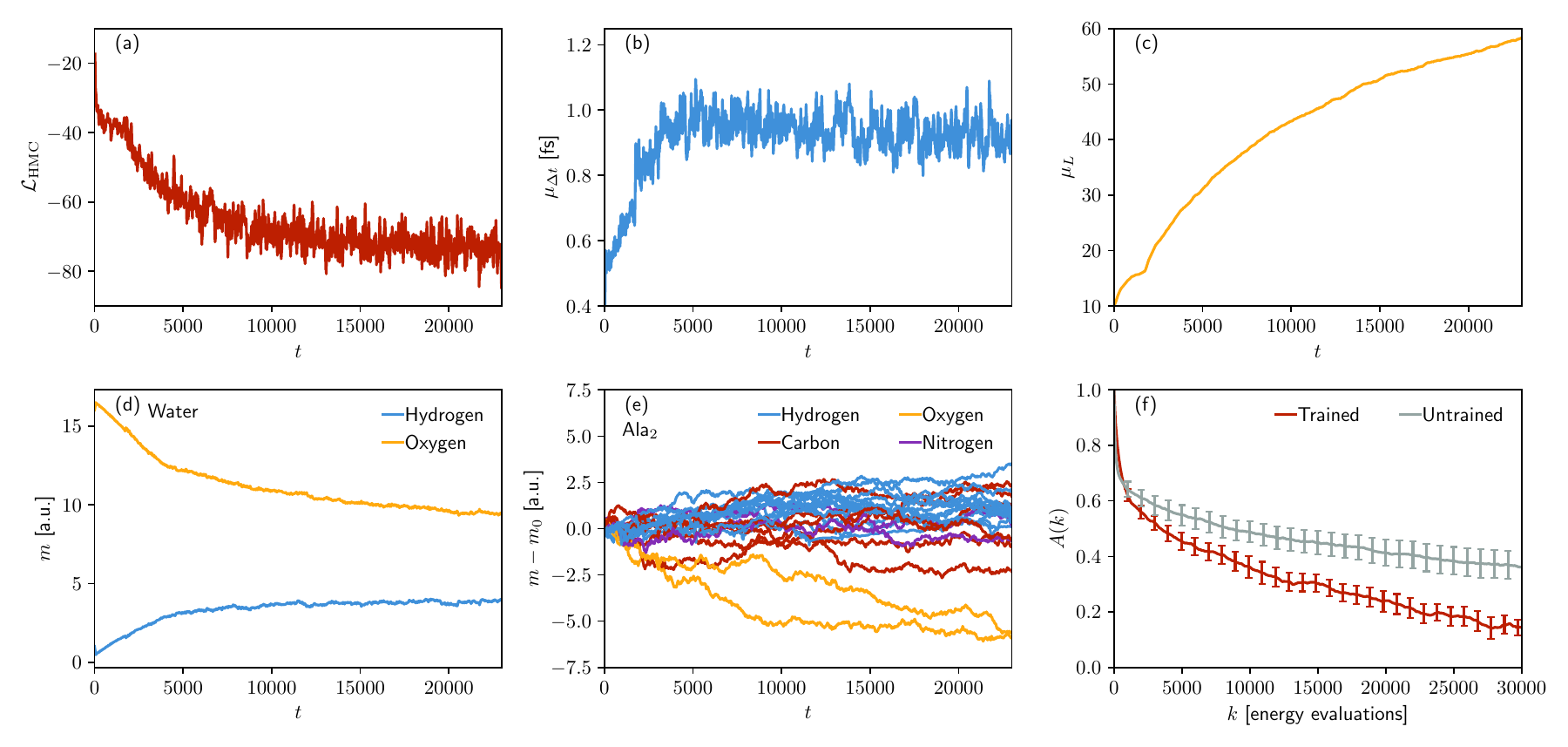}}
  \caption{Plots demonstrating the end-to-end learning of simulation parameters for HMC. (a) Loss $\mathcal{L}_{\mathrm{HMC}}$ as function of training time $t$. (b),(c) The mean of timestep $\mu_{\Delta t}$ and number of integration steps $\mu_L$ against training time $t$. (d),(e) The behavior of the masses $m$ for (d) the hydrogen and oxygen atoms in the water molecules, optimized as shared parameter between all water molecules and (e) the remaining masses in Ala$_2$. In (e) we plot $m-m_0$ to better highlight the relative change of masses for the diverse atom types. (f) Autocorrelation function $A(k)$ of the potential energy $U(x)$ as function of lag-time $k$ measured in terms of the number of energy evaluations.}
  \label{fig:hmc}
\end{figure*}
In DIMOS, a MCMC simulation is set up by defining the chosen proposals, which are then together with the system used to initialize the simulation object:
\begin{center}
\begin{minipage}{\mywidth}
\begin{lstlisting}
hmc_move = dimos.monte_carlo.HMC(T, dt, L, system)
simulation = dimos.MCSimulation(system, positions, [hmc_move], [1.0], T)
\end{lstlisting}
\end{minipage}
\end{center}
In above example, we define a standard HMC move and assign it probability $1.0$ of being picked as proposal distribution when initializing the simulation object.
\par
It is possible to associate several types of MCMC move sets to generate a simulation object, where the different moves are picked with the probabilities specified in the associated list.
Here, we are interested in HMC moves, and to tune its parameters such that the proposal distribution is optimized to reduce the autocorrelation between samples.
For this, we utilize the end-to-end differentiability of DIMOS.
Other packages that in principle could be used are TorchMD\cite{doerr2021torchmd} and TorchSim,\cite{cohen2025torchsimefficientatomisticsimulation} or other Jax or Julia-based simulation engines such as Jax M.D.,\cite{jaxmd2020} FeNNol,\cite{ple2024fennol} or Molly.jl.\cite{Greener2024}
In our original work where we optimized HMC for toy systems and a small protein in vacuum by end-to-end differentation,\cite{christiansen2023self} we indeed relied on TorchMD.\cite{doerr2021torchmd}
However, as presented above, for larger systems this would incur a significant computational overhead.
\par
HMC\cite{duane1987hybrid,neal2011} constructs proposal configurations that are informed by the forces acting on the particles, in contrast to random-walk MCMC where proposals are uninformed.
This allows for large configurational changes without incurring a low acceptance rate.
To obtain a new proposal in HMC, the configuration space is extended to include velocities $v$ (which is usually also done in molecular dynamics simulations).
A proposal is then generated by first sampling velocities $v$ from the Maxwell-Boltzmann distribution, defining the total energy $E(\mathbf{x}, \mathbf{v}) = U(\mathbf{x}) + T(\mathbf{v})$, where $U(\mathbf{x})$ is the potential energy (dependent only on positions $\mathbf{x}$) and $T(\mathbf{v})$ is the kinetic energy (dependent only on velocities $\mathbf{v}$).
In practice, the system is then evolved using a volume-preserving, time-reversible integrator, such as the velocity Verlet integrator of Section~\ref{sec:MD}.
Because phase-space volume of the extended system is conserved by the integrator, the Metropolis-Hastings acceptance criterion does not require computing the Jacobian determinant.
\par
Short trajectories of only a few integration steps $L$ result in diffusive behavior dominated by random velocity resampling.
In contrast, long trajectories in addition to adding additional computational effort risk accumulating numerical errors and redundant exploration (i.e., states ``looping back'').
Similarly, the integrator timestep $\Delta t$ involves a trade-off: Small $\Delta t$ lead to inefficient exploration, while large $\Delta t$ introduces effects from a shadow Hamiltonian or even lead to catastrophic instabilities.
As shown in prior work,\cite{christiansen2023self} naively targeting a fixed acceptance rate can fail to optimize the autocorrelation time under certain conditions.
This is because acceptance rates only reflect energy conservation errors, not exploration quality.
\par
Instead of tuning the acceptance rate, we optimize HMC simulation parameters via end-to-end training, using a physics-informed objective.
Our parameters are the timestep $\Delta t$, the number of integration steps $L$, and the masses of the atoms $m_i$.
We optimize the expected squared jump distance (ESJD)\cite{pasarica2010adaptively} of a system observable $\mathcal{O}$, defined as
\begin{equation}
    \mathcal{L} = \langle (\mathcal{O}'-\mathcal{O})^2 \rangle \approx \frac{1}{R} \sum_{i=1}^{R} \alpha_i (\mathcal{O}'_i-\mathcal{O}_i)^2,
    \label{eq:loss}
\end{equation}
where the acceptance probability
\begin{equation}
\alpha_i = \min(1,\exp(-[E(\mathbf{x},\mathbf{v})-E(\mathbf{x'},\mathbf{v'})]/k_bT))
\end{equation}
depends on the difference of the total energy $E$.
Here, $k_b$ is the Boltzmann factor, $T$ is the temperature, $\mathbf{x}$,$\mathbf{v}$ are the starting position and (randomly picked) velocity, and $\mathbf{x'}$,$\mathbf{v'}$ is the proposed state.
The average is computed over $R$ independent replicas.
For the application in this paper, we set $R=10$ and choose to optimize directly the movement in real space, i.e., we use $\mathcal{O} = x$.
\par
This loss is used as a proxy for the autocorrelation time, defined as
\begin{equation}
A(k) = \frac{\langle \mathcal{O}_t \mathcal{O}_{t+k} \rangle - \langle \mathcal{O}_t\rangle^2}{\langle \mathcal{O}_t^2 \rangle - \langle \mathcal{O}_t\rangle^2},
\end{equation}
where $\langle \ldots \rangle$ symbolizes the thermodynamic expectation and $k$ is the lag-time.
For large $k$, $A(k)$ decays exponentially
\begin{equation}
A(k) \xrightarrow{k \rightarrow \infty} e^{-k/\tau_{\mathrm{exp}}},
\end{equation}
defining the exponential autocorrelation time $\tau_{\mathrm{exp}}$, indicating the number of simulation steps needed to obtain uncorrelated samples.
\par
To optimize $\Delta t$ and the discrete number of steps $L$, we model them as folded normal variables $\mathcal{N}_F(x; \mu, \sigma)$, similar as in Ref.~\onlinecite{errica2023adaptive}, defined on $\mathbb{R}_{>0}$ with density:
\begin{equation}
    \mathcal{N}_F(x; \mu, \sigma) = \frac{1}{\sqrt{2\pi\sigma^2}} e^{-\frac{(x-\mu)^2}{2\sigma^2}} + \frac{1}{\sqrt{2\pi\sigma^2}} e^{-\frac{(x+\mu)^2}{2\sigma^2}}.
\end{equation}
In our case, we learn the mean $\mu$ via backpropagation,\cite{errica2023adaptive,errica2025adaptive} while the variance $\sigma$ is fixed to a relative variance $\sigma_{\Delta t}=0.01\Delta t$ for $\Delta t$ and to the absolute value $\sigma_{L}=1$ for $L$.
In both cases having $\sigma>0$ functions as a way of jittering, commonly used to avoid effects related to suboptimal performance due to the traveled distance being always the same.\cite{christiansen2023self}
\par
To learn an optimized mean $\mu_L$ modelling the length of the trajectory, which is otherwise inherently discrete when considering only the trajectory length $L$ directly, we reweight contributions to the loss $\mathcal{L}$ by the step length $\ell$, yielding
\begin{equation}
    \mathcal{L}_{\mathrm{HMC}} = \frac{1}{R} \sum_{i=1}^{R} \sum_{\ell=1}^{L_q} \frac{w_\ell\alpha_{i,\ell}}{\ell}  (\mathcal{O}'_{i,\ell}-\mathcal{O}_{i,\ell})^2
    \label{eq:loss_steps}
\end{equation}
where $w_\ell = \mathcal{N}_F(\ell; \mu_L, \sigma_L)$, and $\alpha_{i,\ell}$ and $\mathcal{O}_{i,\ell}$ are the acceptance rate respectively system observable to be optimized of replica $i$ at step $\ell$.
The sum is truncated at $L_q$, chosen as the quantile function of $\mathcal{N}_F$ evaluated at $0.999$.
This definition optimizes for computational efficiency in terms of force evaluation, and by approximating the distribution over the length by the quantile avoids infinite sums in the evaluation of the loss.
Note that as final proposal we consider the configuration after $l \sim \mathcal{N}_F(\mu_L, \sigma_L)$ steps, i.e., we sample our length from the discretized distribution describing the weighting of the contribution to the loss.
Alternatively, one could also consider directly a discretized value of $\mu_L$, but would loose the benefits of jittering.
\par
Protein masses are tuned individually, while the water oxygen and hydrogen masses are shared across all water molecules.
This prevents overfitting to changing solvent configurations.
Optimizing the masses is possible in HMC because the partition functions for positions and velocities factorize, hence, the positional distribution is still sampled correctly even though the dynamics does not follow a physical evolution any longer.
Since changes in the \emph{total} mass of the system is trivially equivalent to changes of $\Delta t$, we fix the total mass to be constant and only redistributed masses among the atoms.
\par
Figures~\ref{fig:hmc}(a)-(e) present $\mathcal{L}_{\mathrm{HMC}}$, $\Delta t$, $L$, and the atomic masses $m_i$ during training for a system of Ala$_2$ in explicit solvent modeled by the classical force field AMBER99SB\cite{weiner1981amber} and TIP3P water.\cite{jorgensen1983comparison}
We clearly see that the loss $\mathcal{L}_{\mathrm{HMC}}$ decreases as a function of training time $t$, and reaches a plateau towards the end.
The timestep increases from the small starting value to $\Delta t \approx 1$~fs, which is significantly larger than the $0.5$~fs timestep allowed by the physical dynamics in MD.
In conjunction with the large number of integration steps of $L\approx 60$, this corresponds to big configurational changes for each proposal.
\par
The masses of atoms in the water molecules are optimized for as shared parameters between all molecules.
As shown in Fig.~\ref{fig:hmc}(d), we find that the hydrogen masses are increased while the masses of the heavy oxygen atoms are decreased, similar to what is done manually in MD simulations.\cite{hopkins2015long}
A similar trend can be observed in Fig.~\ref{fig:hmc}(e) for the protein itself, however, with smaller relative changes in the atomic masses.
\par
In Fig.~\ref{fig:hmc}(f) we plot the autocorrelation function $A$ of the potential energy as function of lag time $k$ measured in the number of energy evaluations for a simulation with optimized parameters.
We find that, compared to a simulation with typical parameters of $\Delta t = 0.5$~fs and $L=10$ without mass repartitioning shown in the same plot, the autocorrelations decay significantly faster in the optimized case.
This translates to an exponential autocorrelation time (as measured in computational effort) of $\tau_{\mathrm{exp}} \approx 13,150$ in the trained case versus $\tau_{\mathrm{exp}} \approx 38,900$ in the case where the simulation parameters are chosen ad-hoc, corresponding to a $\approx 3\times$ speed-up in sampling.

\section{Conclusion}
\label{sec:conclusion}
In this work, we have introduced DIMOS, an end-to-end differentiable simulation framework for molecular dynamics and Monte Carlo simulations implemented in \verb|pyTorch|, enabling method development.
DIMOS is designed to be modular and fast, enabling both easy modification and development of novel approaches, and the use as fast simulation engine in particular for MLIPs.
It implements common integrators, barostats, constraint algorithms, and neighborlists, as well as the necessary bonded and nonbonded energy contributions to simulate with classical forcefields.
Common input file formats can be parsed and used to define either completely classical MM, purely MLIP based, or hybrid MM/ML systems.
\par
We have shown that DIMOS has superior performance compared to other fully differentiable simulation frameworks, with a speed-up factor of up to $170\times$.
Ewald and particle-mesh Ewald summation are implemented as advanced approaches to calculate electrostatic interactions, allowing also for the gradient-based optimization of MLIPs based on them.
The mixed precision implementation allows for accurate evolution in time without substantial overhead in computational effort.
\par
Simulations using MLIPs are significantly accelerated using DIMOS when employing high-level simulation engines, with partly larger achievable system size and speed-up of $\approx 1.6\times$ over ASE and about $\approx 5\times$ over TorchSim.
Another strength of DIMOS is demonstrated for three protein systems in explicit solvent, for which we implement a hybrid approach combining classical force fields and MLIPs.
We find that this accelerates the simulations significantly over pure MLIP based modelling.
Additional speed-up is achieved by constraining the movement of fast oscillating atoms, allowing for a $4\times$ increased timestep compared to unconstrained dynamics.
By optimizing end-to-end the parameters of HMC and introducing a data-driven way to perform mass repartitioning, we are able to obtain a $\approx 3\times$ acceleration in sampling efficiency over ad-hoc chosen simulation parameters.
\par
DIMOS is designed to become a vital tool for researchers in computational physics, chemistry, and biology, combining the modularity needed for method development with fast simulation performance.
It is positioned as complementary engine for differentiable method development, and is intended to be used in conjunction with current classical simulation tools.
Due to the modular framework, it is straightforward to add additional features which may be needed in those communities, such as for example the support for more general simulation boxes or virtual particles.
Further performance improvements may be achiveable by implementing multi-GPU support, batching, and direct force calculations for the classical force fields.

\begin{acknowledgments}
We thank Fabio Müller, Falk Selker, Nicolas Weber, and Mathias Niepert for useful discussions.
\end{acknowledgments}

\bibliography{bibtex}

\end{document}